\documentclass[12pt,preprint]{aastex}

\shorttitle{Helical Magnetic Flux Ropes}
\shortauthors{Petrie}

\begin{document}


\title{Potential Magnetic Field around a Helical Flux-rope Current Structure in the Solar Corona}


\author{G.J.D. Petrie}


\affil{National Solar Observatory, 950 N. Cherry Avenue, Tucson, AZ 85719; gpetrie@noao.edu}

\begin{abstract}

We consider the potential magnetic field associated with a helical electric line current flow, idealizing the near-potential coronal field within which a highly localized twisted current structure is embedded.  It is found that this field has a significant axial component off the helical magnetic axis where there is no current flow, such that the flux winds around the axis.  The helical line current field, in including the effects of flux rope writhe, is therefore more topologically complex than straight line and ring current fields sometimes used in solar flux rope models.  The axial flux in magnetic fields around confined current structures may be affected by the writhe of these current structures such that the field twists preferentially with the same handedness as the writhe.  This property of fields around confined current structures with writhe may be relevant to classes of coronal magnetic flux rope, including structures observed to have sigmoidal forms in soft X-rays and prominence magnetic fields.  For example, ``bald patches'' and the associated heating by Parker current sheet dissipation seem likely.  Thus some measurements of flux rope magnetic helicities may derive from external, near-potential fields.  The predicted hemispheric preference for positive and negative magnetic helicities is consistent with observational results for prominences and sigmoids and past theoretical results for flux rope internal fields. 

\end{abstract}



\section{Introduction}

Magnetic fields emerge into the solar atmosphere with significant twist (Leka et al.~(1996). 
Solar coronal eruptions are believed to be caused by the ultimate
failure of the confinement of highly twisted magnetic fields (Low 1996, 2001).  The new twisted flux does not immediately mix with the flux already there, but maintains its identity.  This is because the nearly perfectly conducting plasma and magnetic flux are frozen together except for field reconnection which occurs only on very small scales.  Old and new flux systems find new mutual equilibrium states via successive reconnections and dynamical relaxations.  Taylor's~(1986) theory has described turbulent relaxation to a minimum energy state, a linear force-free field, under the postulated conservation of magnetic helicity contained in the plasma, a postulate shown to be relevant to coronal plasma by Berger~(1984).  In the open environment of the solar atmosphere, the dispersal of magnetic helicity occurs in an unbounded volume so that the ultimate minimum energy state in the absence of further helicity injection is a potential field.  Well-confined structures may maintain their integrity within a near-potential ambient field for long periods.

Examples of confined current structures embedded in relatively current-free surroundings are flux ropes associated with prominences and x-ray sigmoids.  Observations of flux ropes have repeatedly yielded evidence of helical structure (Rust \& Kumar~1996, Chen et al.~1997, 2000, Dere et al.~1999, Wood et al.~1999, Ciaravella et al.~2000, Gary \& Moore~2004).

Observations (Leroy et al.~1984, 1989, Bommier et al.~1994, Lin et al.~1998) have consistently emphasized the presence of strong axial components of prominence magnetic fields.  Amari \& Aly~(1989, 1992) modeled prominences by embedding line currents and current sheets in twisted force-free magnetic fields.  Employing line-current and current sheet solutions, Low \& Hundhausen~(1995) demonstrated the necessity of spatially distributed currents, associated with axial magnetic flux, for prominence support.  Here the mathematical problem can be simplified by assuming that the prominence plasma is cold enough for the width of the plasma condensation to be negligibly thin relative to the scale of the global field configuration.  This is reasonable since the hydrostatic 
scale height at prominence temperatures, a few hundred km, is much smaller than the characteristic length of the large-scale magnetic field, about $10^5$~km.

On the other hand, sigmoids have been identified with plasma structures heated by Parker current sheet dissipation (Titov \& D\'emoulin~1999, Low \& Berger~2003).   Flux surfaces consisting of field trajectories which graze the photosphere, known as bald patches (Titov \& D\'emoulin~1999) are likely locations for the formation of magnetic tangential discontinuities between twisted coronal field and the photosphere.  The two observed sigmoidal shapes, the S-shape and its mirror-reflection, the Z-shape, are each found in both hemispheres of the solar corona, but preferentially in the northern and southern hemispheres respectively.  Associating Z- and S-shaped sigmoids with flux ropes of negative and positive magnetic helicities respectively, Low \& Berger~(2003) found a connection between the writhe that the axial flux in fields around twisted current structures is biased by the writhe of these current structures.  This bias is such that the observed left-handed and right-handed twists of flux ropes in the northern and southern hemispheres may be connected to the hemispheric preferences for the two sigmoidal shapes via the fields around these flux ropes.  Observations of both prominence and sigmoidal flux ropes indicate a hemispheric preference for the handedness of twist of these structures (Canfield \& Pevtsov 1999, Chae~2000, Pevtsov, Canfield \& Latushko~2001).

Titov \& D\'emoulin~(1999) modeled a flux rope using a ring current, with current distributed across a ring of small but finite radius.  An axial field component was produced by threading a line current through the center of this ring.  The field outside the ring is potential.  Low \& Berger~(2003) constructed their flux rope using helically symmetric magnetohydrostatic solutions embedded within an ambient field described by Gold \& Hoyle's~(1960) linear force-free field.  In this paper we shift focus to the field around a localized twisted current-carrying structure: using the potential field around a helical line current we study the relatively unstressed field within which a flux rope with twisted axis (writhe) is embedded.


\section{Magnetohydrostatics and the helical line current}

Consider the static equilibrium of a the magnetic field,
denoted by ${\bf B}$, in the absence of significant plasma forces.
The balance of forces is described by

\begin{eqnarray}
\label{fb}
{\bf j} \times {\bf B}-\nabla p-\rho g{\bf\hat z} & = & 0 ,\\
\label{ampere}
{1 \over 4 \pi} \left(\nabla \times {\bf B}\right) & = & {\bf j},\\
\nabla\cdot{\bf B} & = & 0 , \label{solenoidal}
\end{eqnarray}

Wherever the first term on the left-hand side of Equation~(\ref{fb}) is much larger than the other two terms, as is often the case for the solar corona, the force-free approximation applies:

\begin{equation}
\nabla \times {\bf B}=\alpha {\bf B} .
\label{forcefree}
\end{equation}

\noindent
Here $\alpha$ is a function of space which, by Equation~(\ref{solenoidal}), must be constant along field lines,

\begin{equation}
{\bf B}\cdot\nabla\alpha =0.
\end{equation}

\noindent
Further sub-cases of interest for the study of the corona are the linear force-free case, in which $\alpha =\alpha_0$ is constant and equation~(\ref{forcefree}) becomes linear, and the potential-field (current-free) subcase where this constant $\alpha_0 =0$.  This paper focuses on a special magnetic field which falls into the category of potential field everywhere except a single helical line along which an electric current of infinitesimally thin width flows.  This line current represents a highly twisted flux rope within which all forces of equation~(\ref{fb}) may be important but outside which the field has relaxed to a near-potential state.  Such a field does not explicitly include forces and the line current may be regarded as the limiting case of a class of nonlinear force-free fields whose currents are concentrated along helical axes.

Using Cartesian coordinates $(x,y,z)$ with $y$ in the vertical direction, we define the half-space $y>0$ to be the corona.  Defining the cylindrical coordinates with translational symmetry in the $z$-direction,

\begin{equation}
x=r\cos\varphi,\ y=r\sin\varphi .
\end{equation}

\noindent
Coronal magnetic flux ropes usually lie along relatively straight polarity inversion lines (PILs) at the base of the corona, and may be idealized to be locally 2D with invariance in the direction of the PIL.  Here we are more interested in the global 3D geometry of the ambient field than with the central structure.We define the helical coordinate

\begin{equation}
\zeta=\varphi -kz.
\end{equation}

\noindent
Lines of constant $\zeta$ wind around the cylindrical axis of symmetry $x=y=0$ a full turn over a distance $2\pi /k$ units along that axis.

In a Cartesian 2D field the magnetic surfaces are determined by the conservation of longitudinal magnetic flux and in an axisymmetric field the azimuthal flux is conserved.  The flux surfaces of a helical field are determined by the conservation of the linear combination

\begin{equation}
\psi = \frac{1}{2\pi}(\Psi_z -\Psi_{\varphi}),
\end{equation}

\noindent
of $\varphi$- and $z$-directed fluxes

\begin{equation}
\Psi_z =\int B_z (r,\zeta ) r\ dr\ dz,\,\,\,\,\,\Psi_{\varphi} =\int B_{\varphi} (r,\zeta )\ dr\ dz =\frac{1}{k}\int B_{\varphi} (r,\zeta )\ dr\ d\varphi .
\end{equation}

\noindent
We define this flux function $\psi (r,\zeta )$ as follows:

\begin{eqnarray}
rB_r & = & \frac{\partial\psi}{\partial\zeta} ,\label{fluxfunctionBr}\\
B_{\varphi}-krB_z & = & -\frac{\partial\psi}{\partial r} .\label{fluxfunctionBphiBz}
\end{eqnarray}

\noindent
This solves Equation~(\ref{solenoidal}), fixing the field components in the $\bf r$ and $\nabla\zeta$ directions but leaving the component in the $\bf r\times\nabla\zeta$ direction free.  From the $\varphi$- and $z$-components of Equation~(\ref{fb}) this component $B_z+krB_{\varphi}$ is a free function $F(\psi )$ of the flux function $\psi$ (Neukirch~1999).  Therefore the magnetic field takes the form (Low \& Berger~2003)

\begin{equation}
{\bf B}=\frac{1}{r}\frac{\partial\psi}{\partial\zeta}{\bf\hat r} + \frac{1}{1+k^2r^2}\frac{\partial\psi}{\partial r}{\bf a} + \frac{F(\psi )}{1+k^2r^2}{\bf b} ,
\label{helicalfield}
\end{equation}

\noindent
where

\begin{equation}
{\bf a}=\hat{\mbox{\boldmath $ \varphi $}} -kr{\bf\hat z} ,\,\,\,{\bf b}=kr\hat{\mbox{\boldmath $ \varphi $}} +{\bf\hat z} ,
\end{equation}

\noindent
and the equilibrium of Equation~(\ref{fb}) is described in this case by the Grad-Shafranov equation (Freidberg~1980, Low \& Berger~2003),

\begin{equation}
\frac{1}{r}\frac{\partial}{\partial r}\left(\frac{1}{1+k^2r^2}\frac{\partial\psi}{\partial r}\right) +\frac{1}{r^2}\frac{\partial^2\psi}{\partial\zeta^2} -\frac{2kF}{(1+k^2r^2)^2} +\frac{1}{1+k^2r^2}F\frac{dF}{d\psi} +4\pi\frac{dp}{d\psi}=0.
\label{GradShafranov}
\end{equation}

\noindent
This equation ignores gravity because helical symmetry and gravitational stratification are mutually exclusive.  Here $p$ must be a function of the flux function $\psi$ (Freidberg~1987, Neukirch~1999).  The force-free case is $p\equiv 0$ and the linear force-free and potential cases by $F=\alpha_0\psi$ and $F\equiv 0$.

In the present work we are ignoring plasma forces.  Since we are modeling our magnetic filament using a line current, our model does not resolve the physics of this current.  In real systems the current will have a field-aligned component not associated with a Lorentz force and a further component associated with such a force.  Here, these details are collapsed into a helical line of infinitesimally small width, within which all currents are confined.  The linear force-free form of this Equation~(\ref{GradShafranov}) has eigenfunctions composed of Bessel functions of $r$ and cosines of $\zeta$.  Gold \& Hoyle's (1960) force-free flux rope solution, whose field lines are confined to lines of constant $\zeta $, is the degenerate case $\psi =$ constant with $F$ a non-zero constant.

To find the magnetic field of a single helical line current, we follow Morozov \& Solov'ev~(1966) in first representing the potential fields ${\bf B}={\bf \nabla}\phi^i$ inside and ${\bf B}={\bf \nabla}\phi^e$ outside the cylinder $r=a$ associated with current flowing in the surface of the cylinder with components

\begin{equation}
{\bf j}=i_nka\sin n\zeta\ \hat{\mbox{\boldmath $ \varphi $}} +i_n\cos n\zeta\ {\bf\hat z}
\label{helicalj}
\end{equation}

\noindent
with the potential

\begin{equation}
\label{potentialcases}
\phi (r,\zeta ) = \left\{ \begin{array}{ll}
\phi^i (r,\zeta ) , & \mbox{for $r < a$},\\
\phi^e (r,\zeta ) ,& \mbox{for $r > a$},
\end{array}\right.
\end{equation}

\noindent
where

\begin{equation}
\phi^i=a_n^i I_n(nkr)\sin n\zeta ,\,\,\,\phi^e=a_n^e I_n(nkr)\sin n\zeta .
\label{harmonics}
\end{equation}

\noindent
The boundary conditions at the boundary of the cylinder, derived from Equation~(\ref{ampere}), are

\begin{equation}
B_r^i-B_r^e=0,\,\,\,B_{\varphi}^i-B_{\varphi}^e=-\frac{4\pi}{c}j_z,\,\,\,B_z^i-B_z^e=\frac{4\pi}{c}j_{\varphi} .
\end{equation}

\noindent
Using the identity $I_n^{'}(x)K_n(x)-K_n^{'}(x)I_n(x)=1/x$ the coefficients are found to be

\begin{equation}
a^i_n=\frac{4\pi ka^2i_n}{c} K_n^{'}(nka) ,\,\,\, a^e_n=\frac{4\pi ka^2i_n}{c} I_n^{'}(nka) .
\end{equation}

Now a single helical line current of strength $I$ on the cylindrical surface is described by

\begin{equation}
j_z=\frac{I}{k}\delta (\zeta -\zeta_0 )=\frac{I}{2\pi k}\left( 1+2\sum_{n=1}^{\infty}\cos n(\zeta -\zeta_0 )\right) ,
\end{equation}
\begin{equation}
j_{\varphi}=Ik\delta (\zeta -\zeta_0 )=\frac{Ik}{2\pi}\left( 1+2\sum_{n=1}^{\infty}\cos n(\zeta -\zeta_0 )\right) ,
\end{equation}

The zeroth harmonics are the constant fields

\begin{equation}
\label{zerothharmonics}
{\bf B}_0^i=\frac{2Ik}{c}\ {\bf\hat z} ,\,\,\, {\bf B}_0^e=\frac{2I}{cr}\ \hat{\mbox{\boldmath $ \varphi $}} ,
\end{equation}

\noindent
where the superscripts indicate internal and external solutions as usual, and the full potentials for the interior and exterior fields are found by summing the other harmonics of Equation~(\ref{harmonics}),

\begin{equation}
\phi^i=\frac{2Ik}{c}\left(z+\sum_{n=1}^{\infty} K_n^{'}(nka)I_n(nkr)\sin n(\zeta -\zeta_0 )\right) ,
\end{equation}
\begin{equation}
\phi^e=\frac{2I}{c}\left(\varphi +2k^2\sum_{n=1}^{\infty} I_n^{'}(nka)K_n(nkr)\sin n(\zeta -\zeta_0 )\right) .
\end{equation}

Therefore the magnetic field is given by

\begin{eqnarray}
{\bf B} (r,\zeta) & = & \left( 4k^2a\sum_{n=1}^{\infty} nK_n^{'}(nka)I_n^{'}(nkr)\sin n(\zeta -\zeta_0 )\right) {\bf\hat r}\nonumber\\
& & +\left( \frac{4ka}{r}\sum_{n=1}^{\infty} nK_n^{'}(nka)I_n(nkr)\cos n(\zeta -\zeta_0 )\right) {\bf a}\nonumber\\
& & +2kI\ {\bf\hat z} ,
\label{interiorfield}
\end{eqnarray}

\noindent
inside the cylinder and

\begin{eqnarray}
{\bf B} (r,\zeta) & = & \left( 4k^2a\sum_{n=1}^{\infty} nI_n^{'}(nka)K_n^{'}(nkr)\sin n(\zeta -\zeta_0 )\right) {\bf\hat r}\nonumber\\
& & + \left( \frac{4ka}{r}\sum_{n=1}^{\infty} nI_n^{'}(nka)K_n(nkr)\cos n(\zeta -\zeta_0 )\right) {\bf a}\nonumber\\
& & +2kI\ \hat{\mbox{\boldmath $ \varphi $}} ,
\label{exteriorfield}
\end{eqnarray}

\noindent
outside the cylinder.

\begin{figure*}
\begin{center}
\resizebox{0.39\hsize}{!}{\includegraphics*{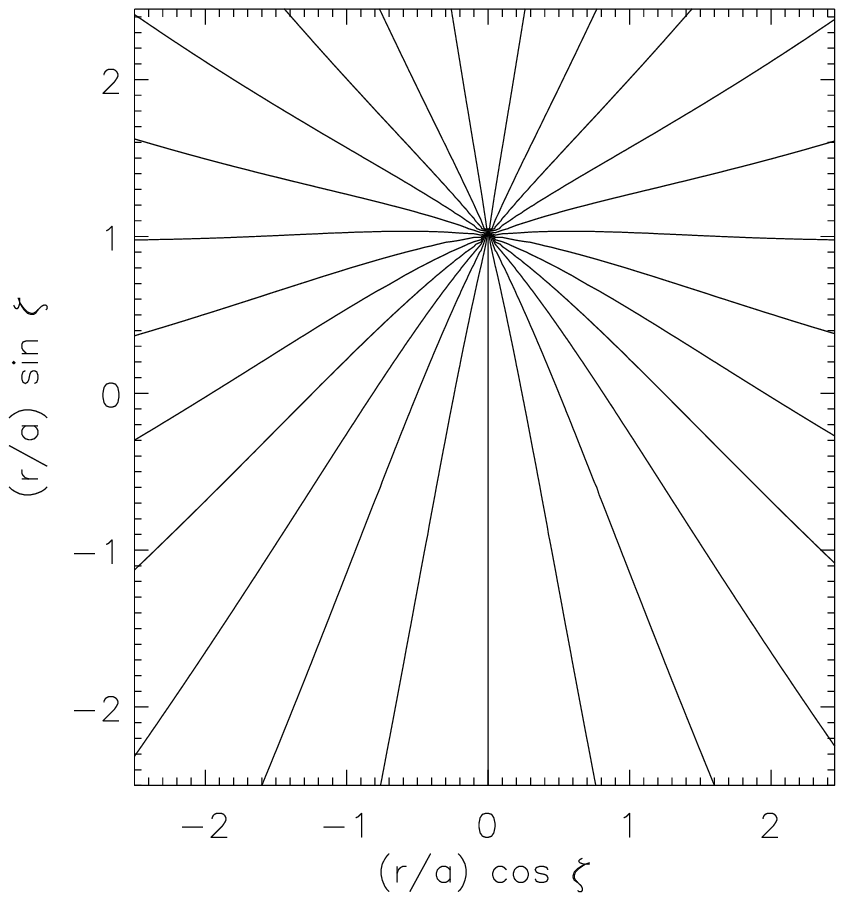}}
\resizebox{0.39\hsize}{!}{\includegraphics*{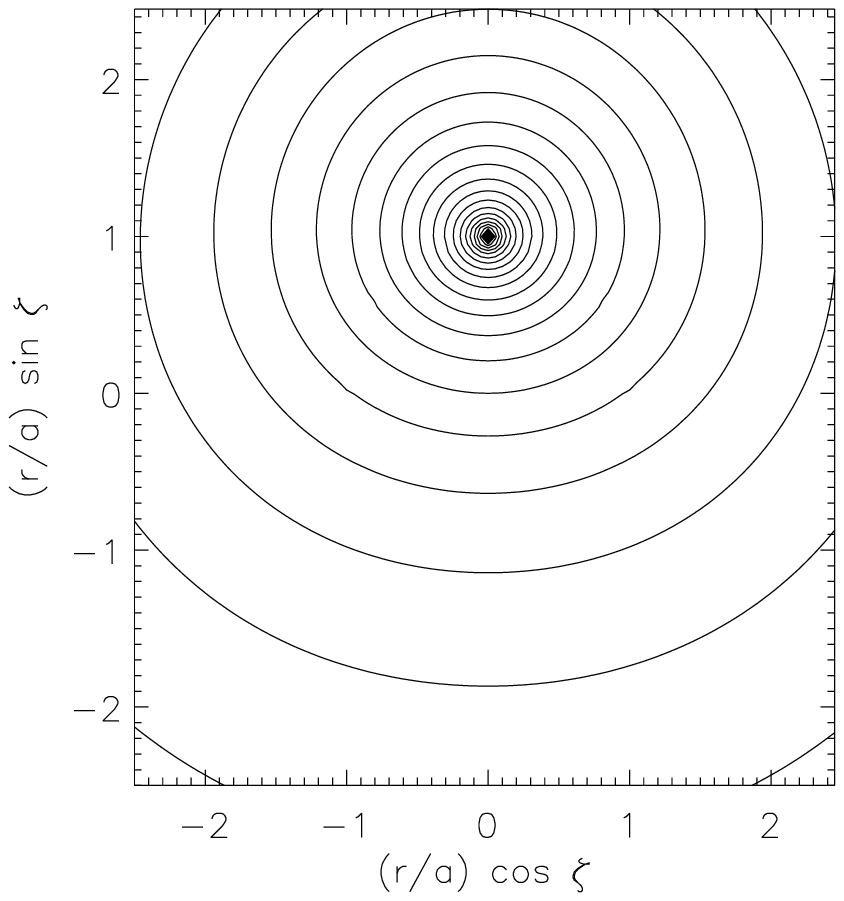}}
\resizebox{0.39\hsize}{!}{\includegraphics*{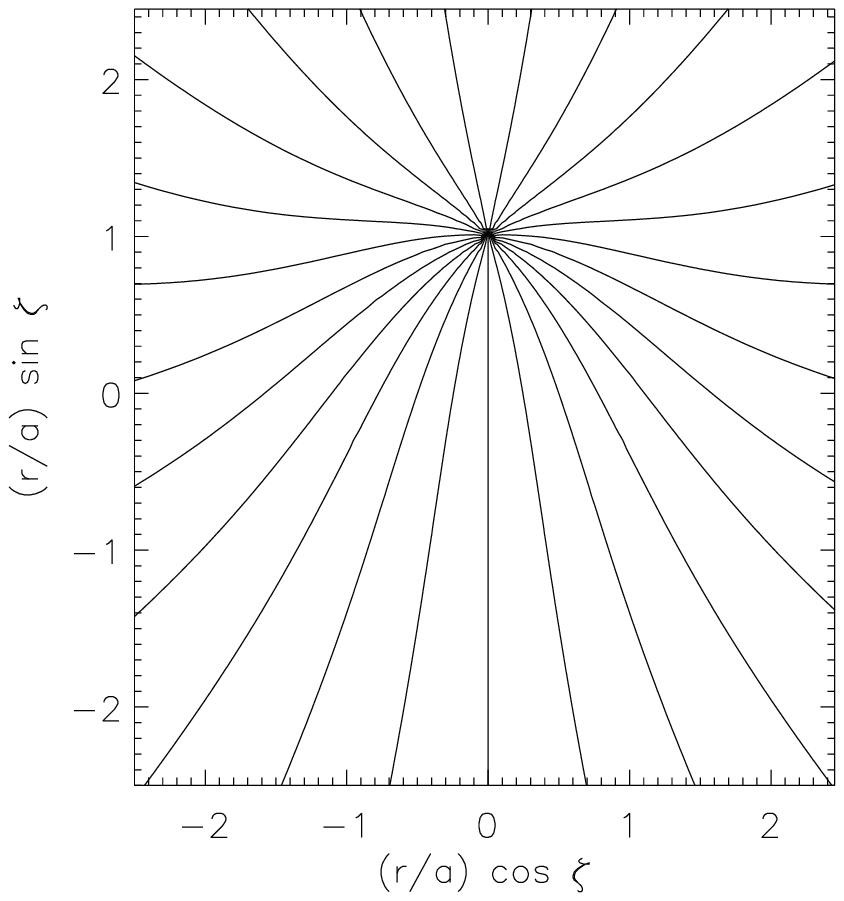}}
\resizebox{0.39\hsize}{!}{\includegraphics*{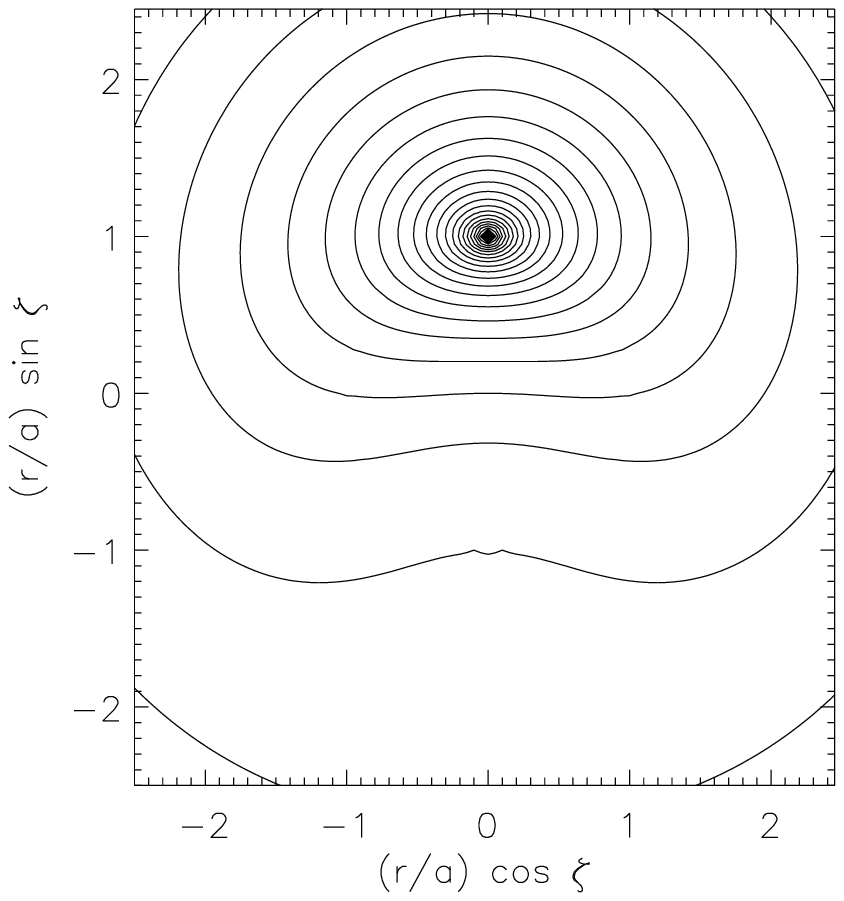}}
\resizebox{0.39\hsize}{!}{\includegraphics*{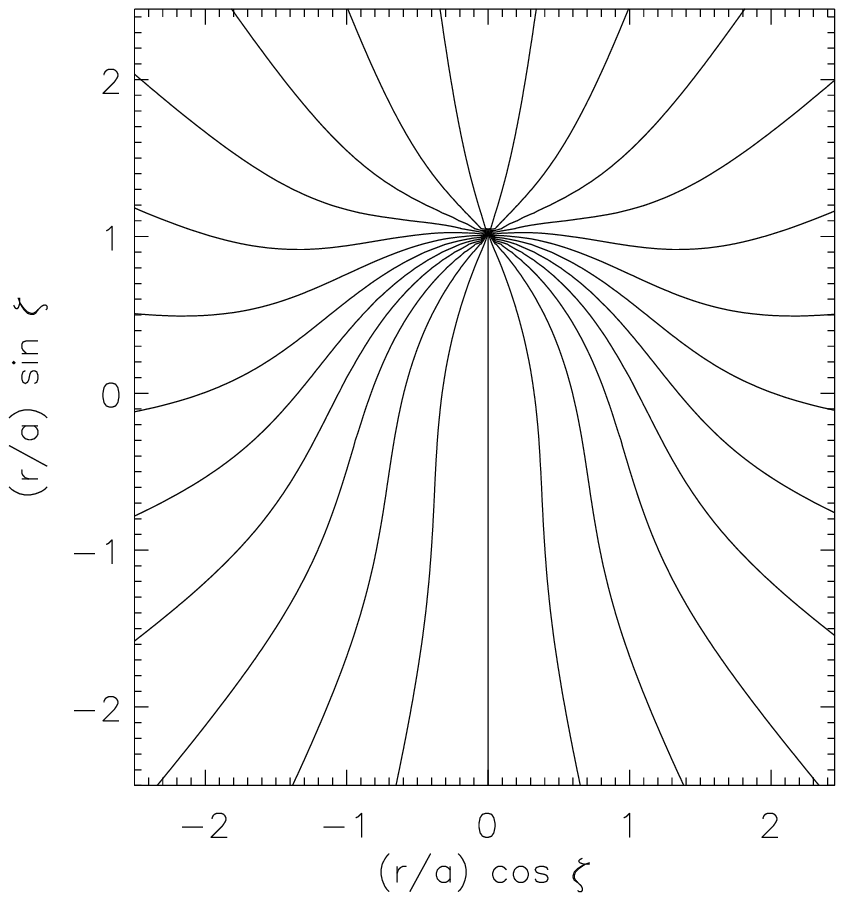}}
\resizebox{0.39\hsize}{!}{\includegraphics*{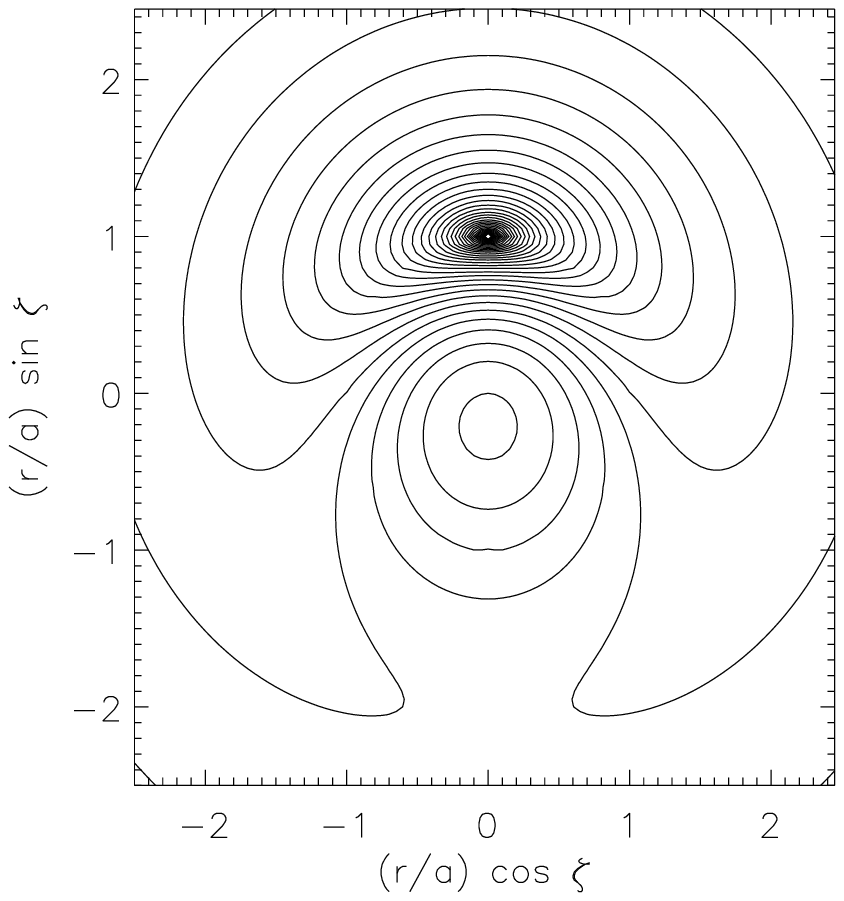}}
\end{center}
\caption{Magnetic potential contours (left pictures) and magnetic flux function contours (right pictures) for the cases $k=0.5$ (top), $k=1.0$ (middle) and $k=2.0$ (bottom).}
\label{fluxcontours}
\end{figure*}

The flux function $\psi (r,\zeta )$ is found in practice from the vector potential $\bf A$, such that $\bf B=\nabla\times A$: $\psi =krA_{\varphi}+A_z$.  This vector potential has been calculated by Tominaka~(2004) and the results are given in the Appendix.  Sample contour plots of both the potential $phi (r,\zeta )$ and the flux function $psi (r,\zeta )$ are shown in Figure~\ref{fluxcontours} for selected values of helical winding number $k$.  Unlike in the 2D Cartesian and axisymmetric cases, the helical flux function does not serve as a matched Euler potential for the magnetic field with the ignorable coordinate, i.e., the vector ${\bf\nabla}\psi\times{\bf\nabla}\zeta$.  Therefore the flux function $\psi$ and the scalar potential $\phi$ do not have contour curves perpendicular to each other everywhere in the $r$-$\zeta$ plane, as Figure~\ref{fluxcontours} clearly shows.  The flux contours are squeezed in the region between the helical axis and the axis $r=0$ because of their relative proximity to the rest of the current and intense magnetic field in the configuration.  This effect increases as $k$ increases until a local extremum of the flux function develops close to $r=0$.  As Morozov \& Solov'ev~(1966) report, the presence of flux function extrema in regions where current flow does not occur.  This contrasts with the 2D Cartesian case where the maximum principle for Laplace's equation forbids closed flux contours in current-free regions.  These second extrema may be regarded as low-lying loops beneath the flux rope.

\begin{figure*}
\begin{center}
\resizebox{0.54\hsize}{!}{\includegraphics*{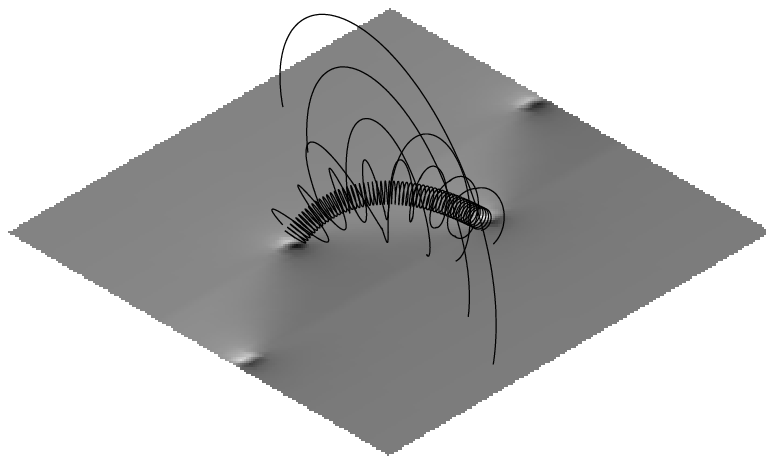}}
\resizebox{0.44\hsize}{!}{\includegraphics*{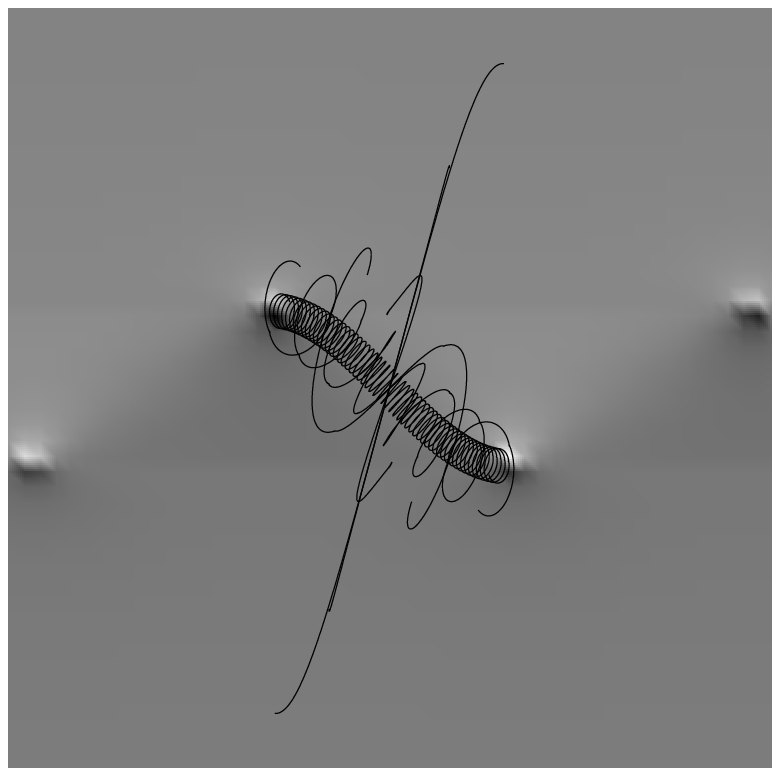}}
\resizebox{0.49\hsize}{!}{\includegraphics*{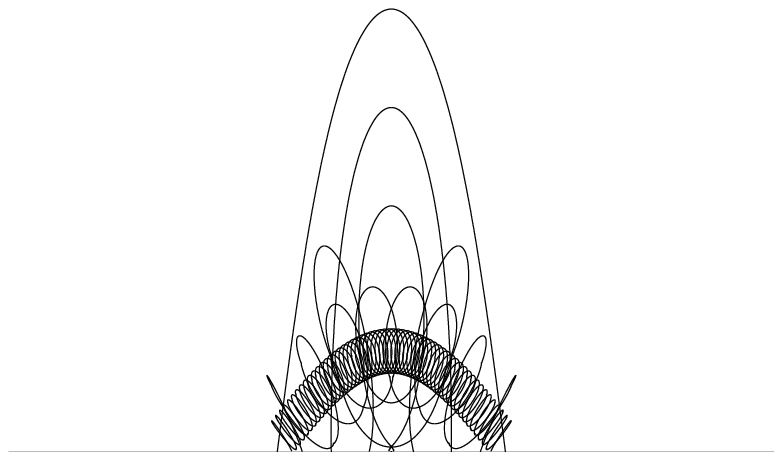}}
\resizebox{0.49\hsize}{!}{\includegraphics*{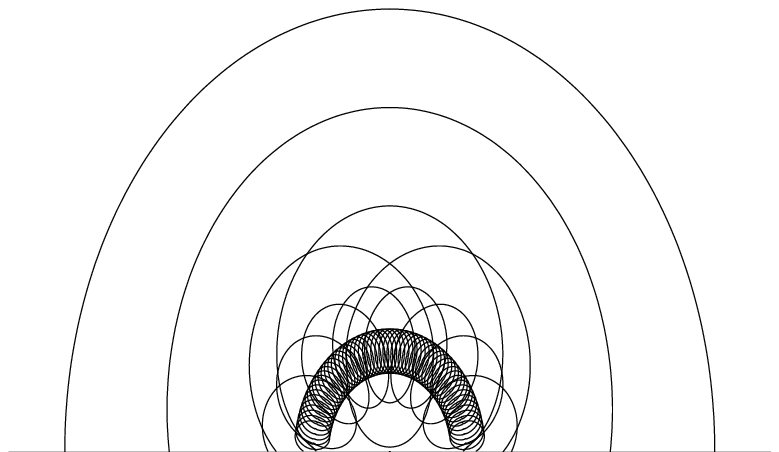}}
\end{center}
\caption{Some sample magnetic flux trajectories for the case $k=1$, viewed from an oblique angle (top left), from above (top right), along the $x$-axis (bottom left) and along the $y$-axis (bottom right).  The shading of the coronal base $y=0$ represents the vertical field component from white, maximum positive flux, to black, maximum negative flux.}
\label{fieldlines}
\end{figure*}

There is a further departure from the simpler 2D Cartesian and axisymmetric cases.  The quantity

\begin{equation}
F = B_z+krB_{\varphi} = 2kI,
\end{equation}

\noindent
is non-zero, which contrasts with Equation~(\ref{helicalfield}) for the potential case $F\equiv 0$.  The non-zero $F$ is entirely due to the zeroth harmonics of Equation~(\ref{zerothharmonics}), which are necessary for a correct description of an infinitesimally thin helical tube of current.  This field has a strong axial ($\bf b$-directed) component, unlike the simpler straight line current and circular ring current solutions (Jackson 1965, Low \& Hundhausen 1995).  In constructing their circular current structure, Titov \& D\'emoulin~(1999) found it necessary to thread a straight line current through their current ring in order to produce an axial magnetic field component.  In contrast, our helical line current naturally produces a significant axial field component because of the skewed curvature of the current trajectory.  Figure~\ref{fieldlines} shows sample field lines for the case $k=1$.  The presence of axially-directed magnetic flux is clear.  The magnetic field structure has the familiar appearance of a twisted flux rope confined within a relatively unsheared arcade.

The existence of strong axial field components has been a strikingly consistent feature of prominence observations (Leroy et al.~1984, 1989, Bommier et al.~1994, Lin et al.~1998).  Low \& Hundhausen~(1995) proved that spatially distributed currents, associated with axially-directed magnetic flux, are a necessary condition for prominence support.  Here we are less concerned with the weight support problem than with relating the axial flux around the flux rope to the global structure of the rope, in particular the writhe.  While these internal and external axial fluxes play separate roles in the physics of the prominence flux rope, they may be coupled in some way.  In real systems it is likely that spatially distributed currents of complex structure exist in the vicinity of the flux rope concentration itself.  This flux could flow in either axial direction producing flux twisting in either direction about the helical axis.  However, the fact that the field surrounding a flux rope with writhe has axial flux even in the absence of local currents implies a preferred direction for axial flux determined by the writhe of the rope.  Thus we expect a preference for right-handed and left-handed twist around right-handed and left-handed flux ropes respectively.

\clearpage
\clearpage

The axial flux of the helical line current potential field may also have a bearing on the hemispheric preference of S-shaped sigmoids and their mirror reflections, which we call Z-shaped after Low \& Berger~(2003).  Z- and S-shaped sigmoids are associated with flux ropes with positive and negative magnetic helicities respectively.  There is no unique relationship between magnetic Z and S shapes on the one hand and the sign of the magnetic helicity on the other because the signs of the writhe of a flux rope and the internal magnetic helicity of the field inside the rope are distinct. Moreover, a flux rope of a fixed helicity can have some field trajectories project into S shapes and others into Z shapes.

The twist of the line current in the example in Figure~\ref{fieldlines} is right-handed.  The twist of the field trajectories around this line current is also right-handed.  This line current is parallel to a Gold-Hoyle magnetic field trajectory.  If the shapes of this line current trajectory projected on the atmospheric base $y=0$ is identified with the shapes of observed sigmoids, this solution implies an association of positive magnetic helicity, that is, a right-handed wind of lines of force, with Z sigmoids.  If we reverse the sign of the entire field, the field preserves its handedness and sign of magnetic helicity because magnetic helicity and handedness are independent of the direction of the magnetic field. If we reverse the sign of the parameter $k$ we obtain the mirror reflection of the solution in Figure~\ref{fieldlines}. In the mirror-reflected solution, the field now has negative helicity and a left-handed wind in its fields. Its line current projected onto $y=0$ would then be shaped like an S sigmoid.

A direct association of the handedness of the line current twist to sigmoidal shape identifies sigmoidal shapes with the writhes of flux ropes.  This is equivalent to Low \& Berger's association of their Gold-Hoyle external fields with sigmoidal patterns since our line currents are parallel to Gold-Hoyle fields. If the observed coronal sigmoids are interpreted in terms of these associations, the implication is that positive and negative helicities are preferentially found in the northern and southern hemispheres.  As Low \& Berger explain, this contradicts the hemispherical preference effect for prominence and sigmoid flux ropes.

Low \& Berger resolved this inconsistency by demonstrating that field trajectories within their flux rope winding around a right-handed axis of helical symmetry with right-handed twist ($\gamma =\alpha_0/k>0$) grazing the atmospheric base project onto the atmospheric base as S-shapes.  Left-twisted lines about a right-twisted axis do not do this so readily.  Anchored field lines rising from the photosphere may wind more than once in the atmosphere before exiting the atmosphere through their second footpoints. Such winding lines may tangentially touch the atmospheric base from above only to return into the atmosphere.  Low \& Berger demonstrated that the heated plasma in a sigmoid might arise from current sheets forming within a flux rope as opposed to deriving from an external field, and that the sigmoidal shape may be related more to the topology of the flux rope internal field than the topology of the external field.  We have seen that the sigmoidal shapes that a Gold-Hoyle field would produce would not conform to the hemispherical preference of sigmoidal shapes described above.  In contrast, a field within a helically symmetric flux rope of right-handed writhe, with right-handed twist about the helical axis is likely to include lines winding about the axis more than once, grazing the photosphere on the way.  Such lines tend to project as S-shapes.

On the other hand, the Gold-Hoyle field has linear field-aligned currents and does not closely approximate an ambient coronal field in a relaxed, near-potential equilibrium.  The magnetic flux trajectories associated with this helical line-current wind more than a full turn about the rope axis in the atmosphere whereas the Gold-Hoyle field is directed parallel to it. For example, there is a particular field line in Figure~\ref{fieldlines} which dips very close to the atmospheric base, almost grazing it just above the origin at the center of the domain.  This line projects as an S-shape and would show up as a bright heated structure.

Such a relationship between sigmoidal shapes and magnetic helicities invokes certain physical properties of the heated plasma.  The force-free field of the domain where our model is meaningful, $y>0$, may be regarded to match discontinuously a distinct field under different physical conditions in $y<0$, since the atmospheric layers between the corona and the photosphere are comparatively very thin.  The flux surface containing all fields grazing the photosphere is tangential to the boundary $y=0$ between a tenuous atmosphere and a dense region below.  Even if the field varies continuously across this boundary, the field immediately above our grazing field is locally unanchored while the field immediately below this grazing field locally threads across $y=0$ and is rigidly anchored by the dense region in $y<0$.  If perturbed, the field above has much more freedom of movement than the field below, and so tangential discontinuities are likely to form and dissipate (Parker~1994) in the neighborhood of the grazing field.  Such a region of grazing field is called a ``bald patch'' by Titov \& D\'emoulin~(1999), who first cited such regions as likely sites of current sheet formation and heating.  Since thermal conduction is much more effective along than across magnetic flux trajectories, thermal conduction heats up all plasmas that are magnetically connected to the heated region. Therefore it is reasonable to interpret the sigmoidal emission patterns as picking out flux trajectories threading threading through bald patches where tangential discontinuities have formed.

Trajectories executing a couple of turns about the axis tend to project as S-shapes.  Therefore, like Low \& Berger's internal field and unlike the Gold-Hoyle field, the potential field around a helical line current is capable of reconciling the hemispheric preferences for magnetic helicity and for sigmoidal shape.Fields both inside and outside current-carrying flux ropes may be contributing to this hemispheric preference.  The helical line current solution leads us to expect that magnetic fields surrounding flux ropes of a given handedness prefer to twist about this current structure with the same handedness.  In reality, a flux rope which has initially acquired twist through the kink instability can readily transfer some of its twist to writhe of the same handedness, and this can create twist of this handedness in the surrounding field.  This makes these fields a viable source of ``bald patches'', Parker current sheets and sigmoidal emission patterns.

\clearpage
\clearpage

\section{Conclusion}

Using a simple solution for the magnetic field associated with a helical line current, we have investigated some consequences of the writhe of coronal magnetic flux ropes for the near-potential magnetic field within which they are embedded.  Magnetostatic solutions in realistic three-dimensional geometry are generally intractable because of the complex structure of the governing equations.  Two simplifications allow us to investigate some consequences of three-dimensional structure.  The idealization of highly twisted, confined magnetic flux ropes as line currents allows us to focus on the relatively current free field around a flux rope without having to treat the full magnetohydrostatic problem.  No solutions are known which combine a current-carrying flux rope field and a potential ambient coronal field.  Such an equilibrium is likely to require a numerical treatment.  Helical symmetry is imposed so that the governing equation becomes the helical Grad-Shafranov equation, an elliptic equation with a linear differential operator whose source solution describes a line current and can be computed in closed analytical form. 

The most striking difference between potential fields around straight line currents and ring currents on the one hand and potential fields around helical line currents on the other is the existence of axially-directed magnetic flux in the helical case.  The helical line current field, in including the effects of flux rope writhe, is therefore more topologically complex than these straight line and ring current fields sometimes used in solar flux rope models.  Since axial magnetic flux is central to the study of coronal flux ropes, this geometry-dependent presence of axial flux may help us to understand some processes behind solar activity.  The two main manifestations of solar activity in the corona, flares and coronal mass ejections, are believed to be related to the evolution of flux ropes associated with prominences and x-ray sigmoids (Low~1996, 2001).

The axial field component of the helical line-current potential field manifests itself as a twist of the potential field about the helical axis whose handedness is the same as that of the axis writhe.  This suggests that near-potential fields around flux ropes with writhe of a given handedness twist preferentially with the same handedness as that writhe.  This may have a bearing on the interpretation of observations of certain classes of coronal magnetic flux rope, including x-ray sigmoids and prominence magnetic fields.  As we have seen, the preference for flux-rope writhe and external magnetic field twist to have the same handedness suggest that ``bald patches'' might tend to occur in such fields, and therefore the associated heating by Parker current sheet dissipation seem likely.  Thus some measurements of flux rope magnetic helicities may derive from external, near-potential fields.  The predicted statistical hemispheric preference for positive and negative magnetic helicities is consistent with observational results for prominences and sigmoids and past theoretical results for flux rope internal fields.

\acknowledgements
I thank the referee for helpful comments.  This work was conducted while the author was a participant in the National Aeronautics and Space Administration (NASA) Postdoctoral Program at Goddard Space Flight Center, and was based at National Solar Observatory, Tucson.

\appendix
\section{Magnetic vector potential of helical line current}

This calculation is given in full by Tominaka~(2004) and we quote the results here.  The magnetic vector potential associated with a known electric line current is given by (Jackson~1965)

\begin{equation}
{\bf A} = \int \frac{{\bf j}\ dS}{|{\bf r-r^{'}}|}
\label{biotsavart}
\end{equation}

\noindent
where $\bf j$ is given by Equations~(\ref{helicalj}), describing an infinitely thin helical current embedded in the cylindrical surface $r=a$, which we denote by $S$.



The three components of $\bf A$ are

\begin{equation}
A_r = \left\{ \begin{array}{ll}
-2kaI\sum_{n=1}^{\infty} [K_{n+1}(nka)I_{n+1}(nkr)-K_{n-1}(nka)I_{n-1}(nkr)]\sin n(\zeta -\zeta_0 ) , & \mbox{for $r \le a$},\\
-2kaI\sum_{n=1}^{\infty} [I_{n+1}(nka)K_{n+1}(nkr)-I_{n-1}(nka)K_{n-1}(nkr)]\sin n(\zeta -\zeta_0 ) ,& \mbox{for $r > a$},
\end{array}\right.
\end{equation}

\begin{equation}
A_{\varphi} = \left\{ \begin{array}{ll}
Ikr+2Ika\sum_{n=1}^{\infty} [K_{n+1}(nka)I_{n+1}(nkr)+K_{n-1}(nka)I_{n-1}(nkr)]\cos n(\zeta -\zeta_0 ) , & \mbox{for $r \le a$},\\
Ik\frac{a^2}{r} +2Ika\sum_{n=1}^{\infty} [I_{n+1}(nka)K_{n+1}(nkr)+I_{n-1}(nka)K_{n-1}(nkr)]\cos n(\zeta -\zeta_0 ) ,& \mbox{for $r > a$},
\end{array}\right.
\end{equation}

\begin{equation}
A_z = \left\{ \begin{array}{ll}
-2\log a+4I\sum_{n=1}^{\infty} K_n(nka)I_n(nkr)\cos n(\zeta -\zeta_0 ) , & \mbox{for $r \le a$},\\
-2\log r+4I\sum_{n=1}^{\infty} I_n(nka)K_n(nkr)\cos n(\zeta -\zeta_0 ) ,& \mbox{for $r > a$}.
\end{array}\right.
\end{equation}






\end{document}